\documentstyle[psfig]{mn}

\def\etal{{\it et al.}}
\def\Ni{\noindent}
\def\ros{{ROSAT}}
\newbox\grsign \setbox\grsign=\hbox{$>$}
\newdimen\grdimen \grdimen=\ht\grsign
\newbox\laxbox \newbox\gaxbox
\setbox\gaxbox=\hbox{\raise.5ex\hbox{$>$}\llap
     {\lower.5ex\hbox{$\sim$}}}\ht1=\grdimen\dp1=0pt
\setbox\laxbox=\hbox{\raise.5ex\hbox{$<$}\llap
     {\lower.5ex\hbox{$\sim$}}}\ht2=\grdimen\dp2=0pt
\def\gax{\mathrel{\copy\gaxbox}}
\def\lax{\mathrel{\copy\laxbox}}
\def\rx{RX\,J1724.0+4114}
\def\rxs{RX\,J1724}

\title[A new high-field polar]{Discovery of the high-field polar 
\rx\thanks{Partly based on observations collected at the German-Spanish 
          Astronomical Centre, Calar Alto, operated by the MPI f\"{u}r 
          Astro\-no\-mie, Heidelberg, jointly with the Spanish National
          Commission for Astronomy.}}

\author[J. Greiner, R. Schwarz and W. Wenzel]
{Jochen Greiner$^1$, Robert Schwarz$^{1}$ and Wolfgang Wenzel$^2$\\
$^1$  Astropysikalisches Institut Potsdam,  An der Sternwarte 16, 
 14482 Potsdam, Germany\\
$^2$  96515 Sonneberg, Hauptstr. 40, Germany}

\date{Accepted 1997 November ??. Received 1997 September 17}
\pagerange{\pageref{firstpage}--\pageref{lastpage}}
\pubyear{1997}

\begin{document}

\maketitle
 
\label{firstpage}

\begin{abstract} 
We report the discovery of a new AM Herculis binary (polar) 
as the optical counterpart of the soft X-ray source \rx\ 
 detected during  the ROSAT all-sky survey. The magnetic nature of this 
$V\sim 17^{\rm m}$ object is confirmed by low-resolution spectroscopy 
showing strong Balmer and  HeII emission lines superimposed on a 
blue continuum which is deeply modulated by cyclotron humps.
The inferred magnetic field strength is $50\pm 4$~MG (or possibly even
$\approx$ 70 MG).
Photometric observations spanning $\sim 3$ years reveal a period 
of 119.9 min, right below the period gap. The morphology
of the optical and X-ray light curves which do not show  eclipses by the 
secondary star, suggests a self-eclipsing geometry. We derive a lower limit
on the distance  of $d \gax 250$ pc.
\end{abstract} 


\section{Introduction}

Cataclysmic variables (CVs) are interacting binary stars consisting 
of a primary white dwarf and a low-mass main-sequence secondary which 
fills its Roche lobe. The common feature of these objects is the accretion
of matter from the secondary onto the white dwarf.  
A detailed description of this class is given by Warner (1995). 
In 1977, Tapia reported the detection of (10\%) polarised emission in 
the nova-like variable AM Herculis indicating that the white dwarf 
in this CV has a strong magnetic field. 
Two decades later a well 
established group of about 60 members form an intriguing subclass of CVs: the 
AM Herculis binaries  or polars. 
The magnetic field strength in these systems as measured by Tapia (1977) and 
his successors using polarimetry, Zeeman and cyclotron spectroscopy ranges 
from 7 to 230 Megagauss, and thus is strong enough to lock the rotation of 
the primary to the orbital motion and channel the accreted matter to a 
small area near one or both magnetic
 poles.  This direct, magnetic accretion leads  to the release of hard and 
soft X-rays as well as (partially) polarised cyclotron radiation. 
Due to their high $F_{\rm X}/F_{\rm opt}$ ratio most of the polars were 
discovered by X-ray missions.
It was in particular the ROSAT all-sky survey (RASS) which more than doubled 
the number
of known systems (Watson (1993), Beuermann \& Burwitz (1995)). 
The source presented here was found in a systematic survey for supersoft X-ray
sources  within the RASS database (Greiner 1996)
which revealed a large number of CVs and single white 
dwarfs. The first polar identified from this sample was V844 Herculis
= RX~J1802.1+1804 (Greiner, Remillard and Motch 1995).   
In this paper we present photometric, spectroscopic
and X-ray observations (summarized in Tab.~\ref{logtab})
which led to the discovery of another AM Herculis system, \rx\ 
(henceforth referred to as \rxs).

\begin{table*}
\centering{
\caption{Log of all observations of \rx .}
\begin{tabular}{llllrrr} 
\hline \noalign{\smallskip}
Telescope & Date & Instrument & Spectral range  & \#$^{(1)}$ & Duration &
     $T_{\rm int}$\\
          &      &            &  & & (h) & (sec) \\
\noalign{\smallskip} \hline \noalign{\smallskip}

ROSAT XRT &90 Aug 24/28& PSPC-C & 0.1--2.4 keV &  320  & 0.36 & 7--32 \\ 
ROSAT XRT &93 Mar 8/11 & PSPC-B & 0.1--2.4 keV &   72  & 1.65  & \\ 
ROSAT XRT &95 Aug 5/8  & HRI    & 0.1--2.4 keV &   90  & 3.58 & \\
\noalign{\smallskip}
Calar Alto 3.5~m &92 Oct 1 & Cassegrain  & 3800--7100 \AA  & 1   & 1 & 3600\\
\noalign{\smallskip}
Sonneberg 0.6~m  &94 Jun 15 & EEV--CCD & R              & 62  & 2.34 & 120\\ 
Sonneberg 0.6~m  &94 Jun 19 & EEV--CCD & R              & 65  & 2.89 & 120\\
Sonneberg 0.6~m  &94 Jul 1  & EEV--CCD & R              & 61  & 2.51 & 120\\
Sonneberg 0.6~m  &94 Jul 2  & EEV--CCD & R              & 46  & 2.06 & 120\\ 
Sonneberg 0.6~m  &94 Jul 11 & EEV--CCD & R              & 50  & 2.47 & 120\\ 
Potsdam 0.7~m &97 Apr 1  & TEK--CCD & WL             & 157 & 5.50  & 120\\ 
Potsdam 0.7~m &97 May 28 & TEK--CCD & WL             & 90  & 3.05 & 120\\ 

      %\noalign{\smallskip} 
      \hline \noalign{\smallskip} 
      \end{tabular}

     \noindent{\Ni\small $^{(1)}$ Number of images for photometry or number of 
                             X-ray counts for ROSAT data.}
    }
    \label{logtab}
   \end{table*}

\section{X-ray observations}

The field of \rxs\ was scanned during the ROSAT all-sky survey for four days 
in August 1990. The source was detected (using the {\sc exsas} 
reduction package provided by MPE Garching; Zimmermann \etal\, 1994)
with a mean count rate of 0.245$\pm$0.005 cts/s 
at a best-fit position of $\alpha_{2000} = 17^{\rm h} 24^{\rm m} 05\fs 4$ and 
$\delta_{2000}$ = 41\degr 14\arcmin 08\arcsec\ (using only channels 25--50
for the position determination). The X-ray emission is 
very soft as indicated by the hardness ratio $HR1 = -0.97\pm0.03$, 
where $HR1$ is defined as (H--S)/(H+S), with H (S) being the 
counts above (below) 0.4 keV.

The X-ray flux shows  100\% modulation with a peak count rate of $\sim$1.5 
cts/s and a pronounced faint-phase where the X-ray flux is practically zero  
(formal count rate of $-0.02\pm 0.10$ cts/s). The RASS light curve folded 
over the photometric 
ephemeris as derived in section 4 is shown together with the light curves 
obtained from further pointed observations with the
PSPC in March 1993 and the HRI in August 1995 (PI: Burwitz) in the three 
upper panels of Fig. \ref{lc}. Although the achieved phase coverage is 
poor in each observation, we can constrain the X-ray bright-phase from 
$\phi\sim 0.5$ to $\phi=0$, 
where $\phi=0$ marks the end of the  bright-phase observed in the optical.
The pointed PSPC observation covers just the end of the  bright-phase  
and the faint-phase intensity is again zero ($0.00094\pm 0.002$ cts/s). 
In the HRI observation \rxs\ is covered over nearly all phases. The mean
intensity during the bright phase (0.02 cts/s) is 
about a factor of 10 lower than during the RASS;
note that due to the soft X-ray spectrum the conversion factor 
between PSPC and HRI count rates is 7.5 (Greiner \etal\, 1996). 

\begin{figure}
  \psfig{figure=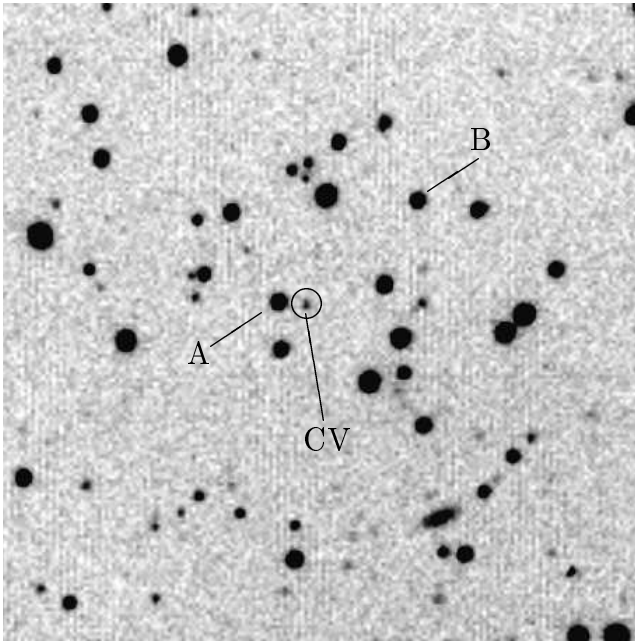,width=\columnwidth,%
    bbllx=125pt,bblly=355pt,bburx=338pt,bbury=571pt,clip=}
  \caption[chart]{
    $R$-band CCD image of \rx\ obtained with the Sonneberg 60 cm telescope. 
    North is top and East to the left. The size of the field is 
    $\approx$ 8\arcmin $\times$ 8\arcmin.
   The uncertainty of the X-ray position is given by the 95\% confidence error
   circle (10\arcsec) which is based on the HRI observation.
   The position of the optical counterpart
   is $\alpha_{2000} = 17^{\rm h}24^{\rm m}06\fs 2$ and $\delta_{2000}=41\degr
   44\arcmin 09\arcsec$ ($\pm 1 \arcsec$). Star `A' is used for differential 
   photometry (Fig. \ref{lc}) and star `B' for the evaluation of the
   long-term behaviour on the Sonneberg astrographic patrol plates 
   (paragr. 5). }
   \label{chart}
\end{figure}

The HRI observation allows for an improved X-ray position ($\pm 10\arcsec$) 
determination as compared to the original RASS error circle 
(see Fig. \ref{chart}):
$\alpha_{2000} = 17^{\rm h} 24^{\rm m} 06\fs 2$,
$\delta_{2000} = 41\degr 14\arcmin 11\arcsec$.
This confirms the earlier (1992) identification of \rxs\ 
based on the optical spectroscopy.

Only the RASS data are suited for a spectral investigation since the other
PSPC observation only covers the faint phase interval (Fig. \ref{lc}).
Adopting a blackbody plus thermal bremsstrahlung model and fixing
the temperature of the latter component to 20 keV gives the values
reported in Tab. \ref{fit}. The temperature of the blackbody component
is about 50 eV and the absorbing column in the free fit 
(see Fig. \ref{oxspec}) 
about 2/3 of the total galactic absorbing column in this direction 
(Dickey and Lockmann 1990). 
We also have performed fits with fixing either the absorbing column to the 
galactic value (2.7$\times$10$^{20}$ cm$^{-2}$) or the blackbody temperature
to the canonical value of 25 eV. In both cases the fit is considerably 
worse.

\begin{table}
\caption{Results of three different spectral fits of RASS data of \rx\ with 
the sum of a blackbody and thermal bremsstrahlung model. Parameters marked
with a $^*$ have been fixed during the respective fit: the second line for
adopting the galactic absorbing column and the third line for 
fixing the blackbody temperature to the canonical value of polars.}
\begin{tabular}{ccccc}
\hline \noalign{\smallskip}
                      & \multicolumn{2}{c}{blackbody } & bremsstrahlung  \\
     $N_{\rm H}$      & k$T$   &  Norm       & Norm  \\
     (10$^{20}$ cm$^{-2}$)     & (eV) & (ph/cm$^2$/s/keV) 
                                   &  (ph/cm$^2$/s/keV) \\
\noalign{\smallskip} \hline \noalign{\smallskip}
    1.9$\pm$0.5  &  58$\pm$20 & 1.4$\times$10$^{-2}$ & 1.9$\times$10$^{-6}$ \\
    2.7$^*$      &  51$\pm$20 & 2.8$\times$10$^{-2}$ & 6.5$\times$10$^{-6}$ \\
    5.6$\pm$0.5  &  25$^*$    & 240$\times$10$^{-2}$ & 20.5$\times$10$^{-6}$ \\
    \hline 
     \end{tabular}
     \label{fit}
      \end{table}

With the parameters of the free fit and 
considering only the X-ray bright phase, the unabsorbed fluxes of the two 
model components in the ROSAT band (0.1--2.4 keV) are
$F_{\rm bbdy}$ = 1.3$\times$10$^{-11}$ erg/cm$^2$/s and
$F_{\rm thbr}$ = 2.8$\times$10$^{-14}$ erg/cm$^2$/s, giving a flux ratio of
$F_{\rm thbr}$/$F_{\rm bbdy}$ = 0.0022. The mean luminosity 
(of both components) during the X-ray bright phase is 
$L_{\rm X}$ = 1.5$\times$10$^{31}$ (D / 100 pc)$^2$ erg/s.

\section{A low resolution spectrum}

\begin{figure*}
  \psfig{figure=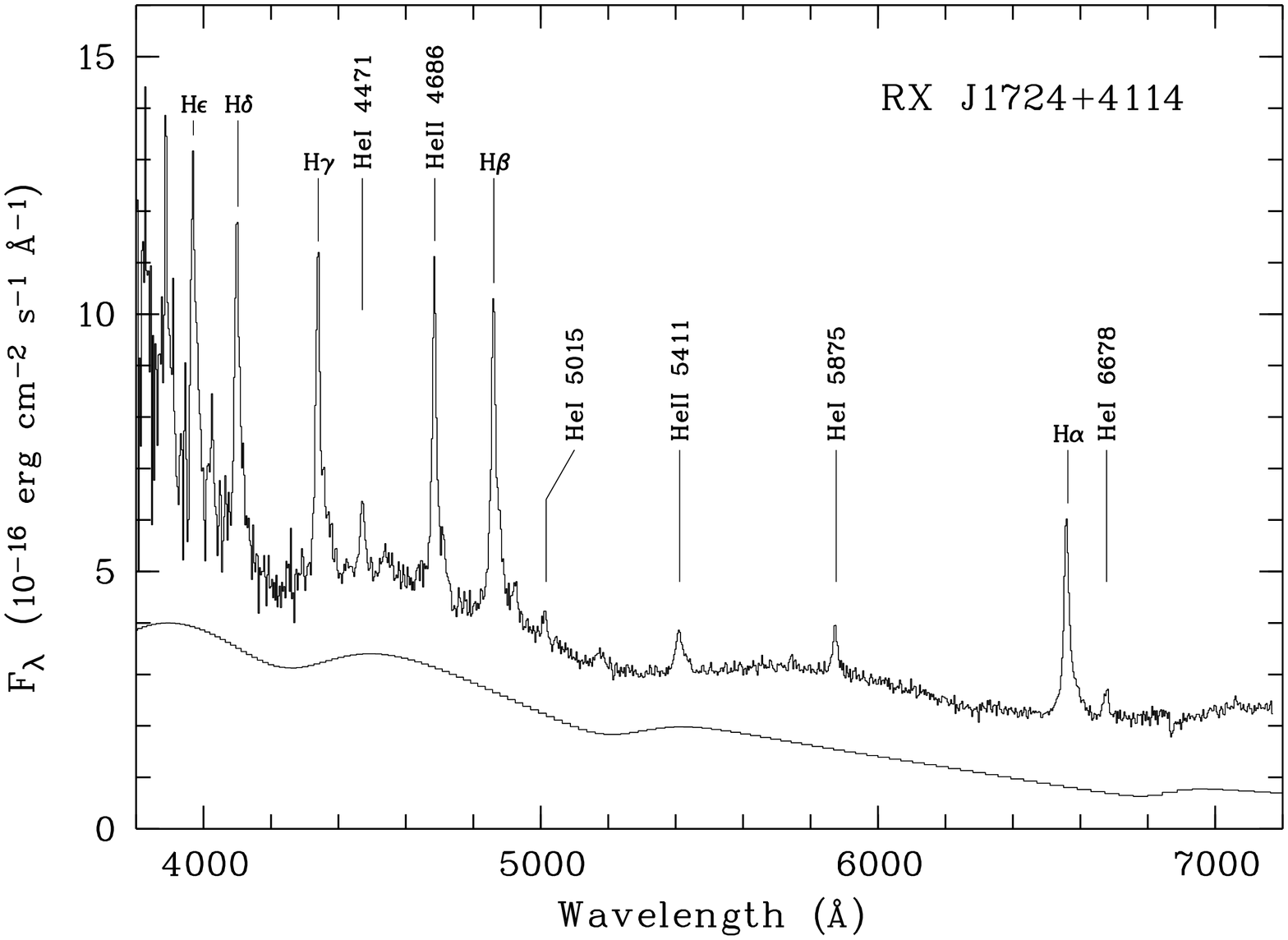,width=11.7cm,%
      bbllx=565pt,bblly=67pt,bburx=45pt,bbury=800pt,angle=-90,clip=}
  \vspace*{-8.2cm}\hspace*{11.4cm}
  \psfig{figure=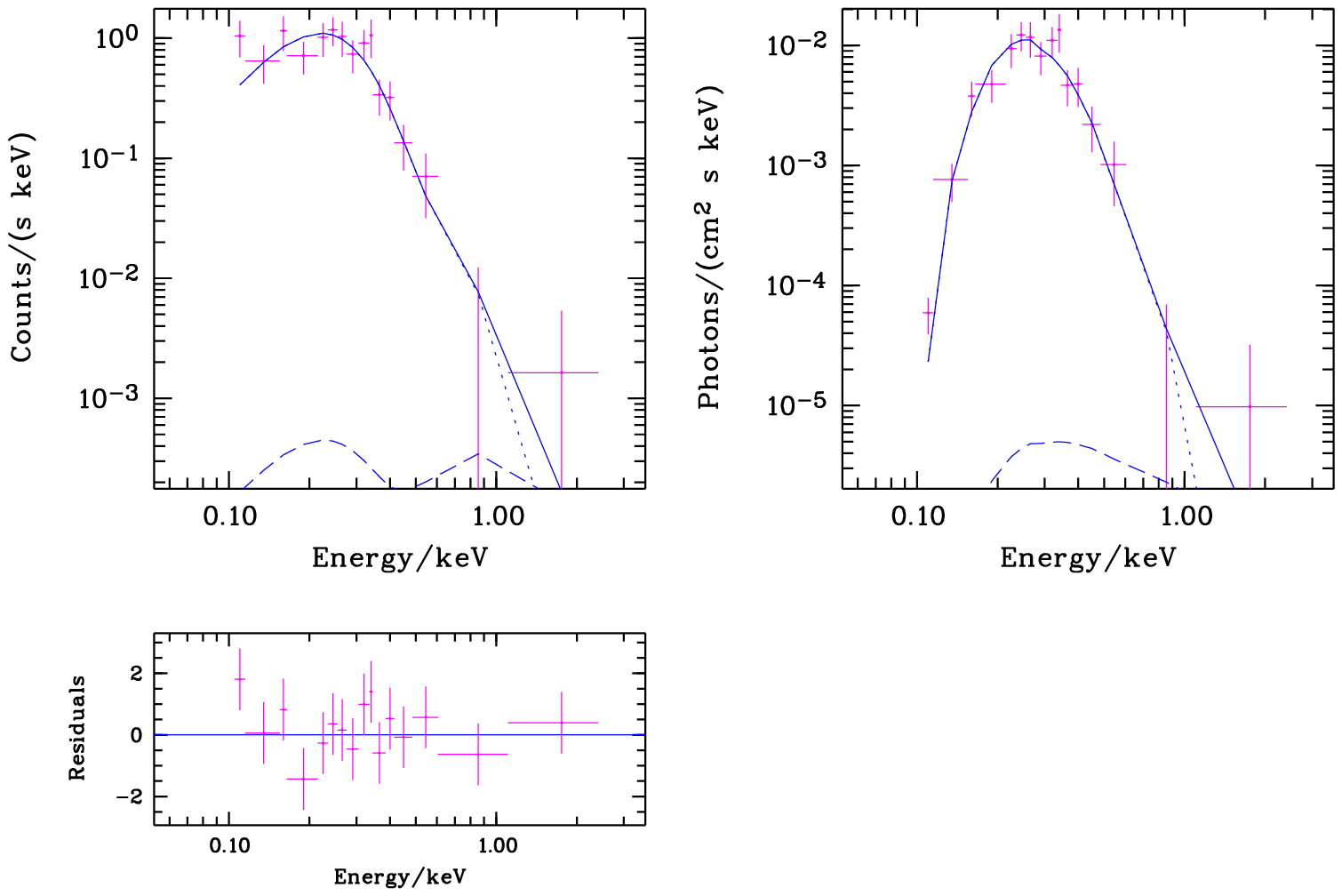,width=6cm,%
       bbllx=2.5cm,bblly=1.3cm,bburx=10.cm,bbury=11.6cm,clip=}
  \caption[]{{\bf Left:} Low resolution optical spectrum of RX~J1724 obtained 
      on October 1, 1992. Main emission lines are indicated. The solid line 
      represents a homogeneous cyclotron model for a magnetic field of 50 MG 
      assuming $kT=20$ keV, $\log \Lambda=3.5$ and $\theta =70\degr$.
      {\bf Right:} Phase-averaged RASS X-ray spectrum of RX~J1724 unfolded 
      with a blackbody
      plus thermal bremsstrahlung spectrum (see text for details). The lower
      right panel shows the residua of the fit in units of $\sigma$.}
  \label{oxspec}
\end{figure*}

Within the program of the optical identification of all new ROSAT
supersoft X-ray sources we obtained a low resolution (FWHM $\sim 12$ \AA) 
spectrum of \rxs\
on October 1, 1992 with the  3.5~m telescope at Calar Alto, Spain. We used
the Cassegrain spectrograph equipped with a RCA CCD as detector
covering the optical wavelength range from  3800--7100 \AA. The
observation was obtained   under stable photometric conditions and
accompanied by measurements of the standard star Feige~110 which was used
to calibrate the flux with an accuracy of $\sim 20$\% (using standard
{\sc midas} procedures). By convolving the
original spectrum with functions representing the $BV\!R$ bandpasses we arrive 
at $V=17\fm 6$, $R=17\fm 1$ and $B=17\fm 6$ for \rxs.

The exposure
lasted one hour corresponding to approximately half of the binary orbit, and
was centered on HJD 244\, 8897.30923 which corresponds to $
\phi = 0.619$ of the ephemeris given by Eq.~2. Hence the spectrum covers
equally $\sim 50$ \% of the faint and bright phase.
The original spectrum is shown in Fig. \ref{oxspec}. 
It is dominated by intense
emission lines of the Balmer series, He\, {\sc II} $\lambda 4686$\AA,
and He\, {\sc I} superimposed on a blue continuum. The inverted Balmer
decrement and the strength of the  He\, {\sc II} $\lambda 4686$\AA\
line point to a magnetic CV classification.  

This is directly confirmed by the cyclotron lines seen at $\lambda 4500$
and $\lambda 5700$. According to

\begin{equation}
\lambda_{n} = \frac{10710}{n}\left( \frac{10^{8}}{B {\rm(G)}}\right) ~~ {\rm \AA}
\end{equation}

\noindent  the separation of the two observed cyclotron
humps allows for an interpretation as the 3rd/4th or 2nd/3rd harmonics of
the cyclotron fundamental.
If we identify these as the 3rd and 4th harmonics, the (minimum) implied field 
strength is   50$\pm$4 MG
depending on the plasma temperature and the viewing angle. A 
corresponding cyclotron model for a 20 keV plasma, a polar angle of 70\degr\, 
and a plasma parameter $\log \Lambda = 3.5 $  is  shown in Fig. \ref{oxspec}. 
The high polar angle is thought to be a reasonable assumption due to the
accretion geometry (see below) and the fact that the exposure of the 
spectrum covers the rise of the spot over the limb. 
In the case of an interpretation of the observed cyclotron humps  as the 2nd 
and 3rd harmonics the implied field strength is of the order of 70 MG.
Further low-resolution spectroscopy extending to the near-infrared is
needed to clarify this ambiguity.

\section{Photometric observations}

 \begin{table}
 \caption{Heliocentric Julian dates of the end of the bright phase 
(estimated error of 180\,s) together with 
the epoch and deviation from phase zero implied by the ephemeris of Eq.~2.}
  \begin{tabular}{lrrl}
\hline \noalign{\smallskip}
  \multicolumn{1}{c}{$T_{\rm end}$} &  \multicolumn{1}{c}{O-C} &\multicolumn{1}{c}{Epoch} & Filter \\
  \multicolumn{1}{c}{(HJD 2400000+)} & \multicolumn{1}{c}{(sec)} & & \\ 
    \noalign{\smallskip} \hline \noalign{\smallskip}
       49519.4735 &  72 &       0 & R  \\
       49523.4690 & 21  &      48 & R  \\
       49536.4615 & 49  &     204 & R  \\
       49545.4567 & -28 &     312 & R  \\
       50540.4467 & -217&   12259 & WL \\
       50540.5325 & 108 &   12260 & WL \\
       50540.6153 & -163&   12261 & WL \\
       50597.4158 & 150 &   12943 & WL \\
       50597.5010 & 86  &   12944 & WL \\
     \hline 
     \end{tabular}
     \label{minima}
      \end{table}

\begin{figure}  
  \psfig{figure=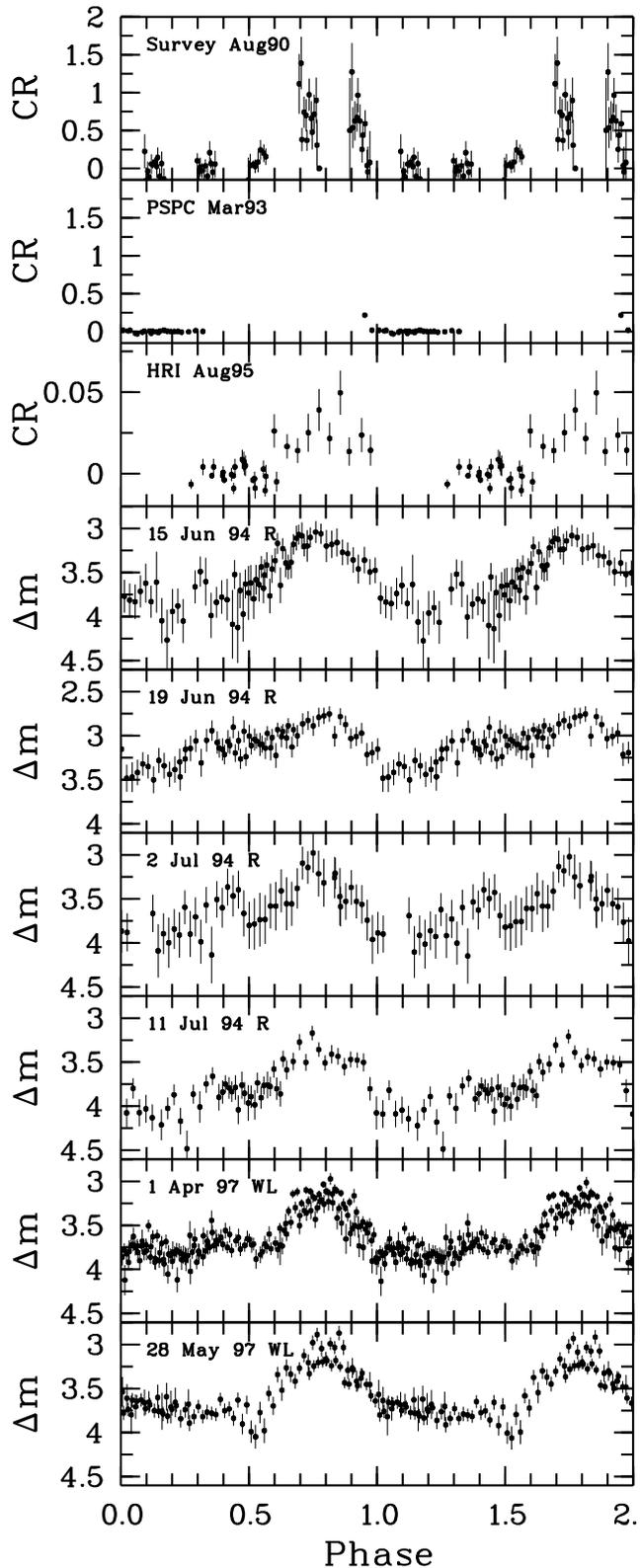,width=\columnwidth,%     2column version
      bbllx=60pt,bblly=68pt,bburx=342pt,bbury=777pt,clip=}
   \caption[Fig3]{
   Optical and X-ray light curves of \rxs\ plotted as a function of ephemeris
   Eq.~2. Data are plotted twice for clarity and units are cts/s in the 
   corresponding X-ray detector for the upper three panels and relative 
   magnitudes with respect to star A (Fig. \ref{chart}) elsewhere.}
  \label{lc}
\end{figure}

\begin{figure}  
  \psfig{figure=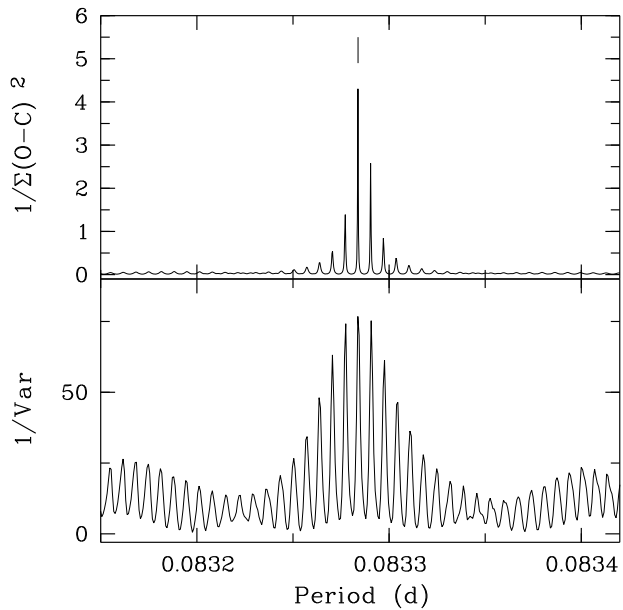,width=88mm,%
      bbllx=560pt,bblly=68pt,bburx=49pt,bbury=600pt,angle=-90,clip=}
  \caption[perio]{\label{perio}  
     Results of the period-search using the $\chi^2$ calculation of the minima 
     given in Tab. \ref{minima} (upper panel, units of the ordinate are 
     $10^{4}$d$^{-2}$) and the analysis-of-variance 
     (lower panel). The adopted period of is marked with a tick.}
\end{figure}

\rxs\ was photometrically monitored during
7 nights in 1994 and 1997 with the 0.6~m reflector at Sonneberg Observatory
and the 0.7~m reflector at the Astrophysical Institute Potsdam (both Germany).
Observational details are listed in Tab. \ref{logtab}.
During all runs no standard stars were observed, so that we are 
restricted to differential photometry which was computed with
respect to star 'A' marked in Fig. \ref{chart}. 
We used the profile-fitting scheme of the {\sc DoPhot} reduction package 
(Mateo \& Schechter 1989) to achieve high accuracy.
The individual light curves  are characterised by a faint phase showing only
little variability followed by a pronounced  
bright phase with an amplitude of 0.7 mag. 
In order to derive an ephemeris we carried out a period search using the 
analysis-of-variance method (Schwarzenberg-Cerny 1989) and a least
squares calculation applied to the heliocentric timings defining the 
end of the  bright phase as compiled in Tab. \ref{minima}. 
The resulting periodograms (Fig. \ref{perio})
reveal a  period of 0\fd 0832843. 
Cycle aliases caused by the $\sim$ 2.8 year 
separation of the data sets are quite prominent. 
However, they lead to a significant phase displacement 
of the observation on May 28, 1997. Moreover, for both neighbouring 
alias periods the X-ray bright and faint phases as observed in  the three 
X-ray observations do not align. We therefore are confident that the alias
periods can be ruled out. Then,
the accuracy is sufficient to connect all the data from August 1990 to May 1997
with an uncertainty of $\delta \phi \sim$ 0.029.
The linear ephemeris for the times of end bright phase derived from the 
optical data is

\begin{equation}
T_{\rm end}({\rm HJD})\,  = 244\, 9519.4721(14) + 0.08328388(8)E,
\end{equation}

\noindent where the numbers in brackets give the uncertainty in the last 
digits.
In Fig. \ref{lc} we present a collection of light curves folded 
over the ephemeris given by Eq.~2. The bright phases observed in the 
optical and X-ray bands coincide.

\section{Discussion and Conclusions}

We have identified the X-ray source \rxs\ as a new magnetic CV.
The combination of extreme soft X-ray spectrum, a strong magnetic field of 
B = 50$\pm$4 MG (or possibly even $\approx$70 MG), an inverted Balmer 
decrement and strong He\,II lines 
provide strong evidence for the polar nature though no polarimetry has 
been obtained. The observed values for $B$ and 
$F_{\rm thbr}$/$F_{\rm bbdy}$ obey well the proportionality relation
between these quantities found for other polars (Beuermann \& Burwitz 1995).

The period derived from the optical and X-ray light curves is 119.9 min,
right below the lower edge of the period gap. No other periodicities
have been found, so that \rxs\ appears to be synchronised over the observed
timescale. 
 
The coincidence in phase of the bright phase in X-ray and optical bands as
well as the lack of X-ray emission during the optical faint phase resemble 
the behaviour of self-eclipsing 
polars like ST LMi or VV Pup (Cropper \& Warner (1986), Cropper (1986)) where
the accretion region passes behind the limb of the white dwarf and is out 
of sight. The duration of the faint phase $\gamma$ gives a constraint 
on possible geometries. Assuming a point-like accretion spot the inclination 
$i$ and the colatitude $\beta$ are related  via 

\begin{equation}
\cos \gamma \pi = \cot i \cot \beta.
\end{equation}

\noindent The lack of eclipses implies $i < 78\degr$.
The duration of the bright phase is $\le0.5$ for most of the observations
consistent with $\beta > 90\degr$, i.e.  an accretion region located on 
the hemisphere of the
white dwarf facing away from the observer (``southern hemisphere''). 
The light curves obtained on June 19
and July 2 1994 show a much more extended bright phase. This might  indicate 
that the location of the accretion spot might have changed or a second 
accreting pole was active. 

The observed X-ray intensity during the HRI pointing in 1995 is considerably 
smaller than the intensity seen during the RASS 5 years earlier, thus implying
variable mass transfer to the white dwarf. We note, however, that the observed
count rate is an extremely sensitive function of the temperature: depending 
on the exact absolute temperature and the model used, the count rate is 
proportional to $T^{5-16}$ (Heise \etal\ 1994). Thus, a reduction of the 
temperature $T$ by about 30\% can account already for the observed count rate
difference. It is therefore impossible to quantify the difference in mass 
transfer rates.

Additional evidence for changes in the transfer rate comes from the
discovery of substantial ($>1$ mag) long-term optical variations using 
photographic patrol plates. \rxs\ is covered by the M\,92 field (taken with
the GB 40/190 cm instrument) of the
Sonneberg Observatory astrographic patrol (though very near the edge of 
the field of view). A check of the available $\approx$100 plates reveals
\rxs\ sometimes (e.g. 1976 Apr. until Sep., 1987 June 30, 1993 Aug. 13--15)
as bright as comparison star B (labeled in Fig. \ref{chart}; note that due to 
its colour star B is notably fainter than star A in the blue band).
At other times it is invisible even on very deep plates (fainter than
17.5 m$_{\rm pg}$) such as 1982 Apr./May,
1983 Jun. 6/7, 1984--1985 and 1992 (including the time of the spectroscopic
observation). Unfortunately, the data are too
spotty to derive a meaningful lightcurve. Due to its clear variability
this object is assigned the number S 10946 in the series of variable stars 
detected at Sonneberg Observatory.

The density of a Roche-lobe filling secondary with \mbox{$P=2$~h} is 
$28\pm 0.5$ g~cm$^{-3}$, only weakly depending on the mass ratio $q$.
Assuming that a mass-radius relationship for main-sequence stars is valid
for \rxs\ (e.g. Patterson 1984) we find $M_{2}=0.16$ M$_{\odot}$ 
and $R_{2}=0.2 $ R$_{\odot}$. The spectral type of a star with that mass
is M~4--4.5 (Kirkpatrick \& McCarthy 1994).
We do not see any spectral signature of the late-type companion in our 
spectrum.  Assuming a contribution of the secondary to the total mean optical
light of $\lax$10\% in the $V$ band ($V \gax 20\fm 1$) and using 
$M_{\rm V}$=13.1 mag (Kirkpatrick \& McCarthy 1994) we
find a  lower limit on the distance of \rxs\ of $d \gax$ 250 pc. 
This is consistent with the $N_{\rm H}$ found in the X-ray spectral 
fit and the galactic latitude of \rxs\ of b$^{\rm II}$ = 33\fdg3.

\section*{Acknowledgements}

RS is very grateful  to Axel D. Schwope for generously providing help and 
software for the cyclotron spectroscopy. 
We thank H. Steinle for assisting the observation on Oct. 1, 1992,
J. Schmoll for that on May 28, 1997 and P. Kroll and R. Luthardt for help at 
the Sonneberg Observatory equipment. We also greatly acknowledge the 
collaboration with V. Burwitz on the 1995 ROSAT HRI observation.
JG and RS were supported by the Deutsche Agentur f\"ur
Raumfahrtangelegenheiten (DARA) GmbH under contract FKZ 50 QQ 9602 3 and
50 OR 9206 8, respectively and WW under contract numbers 05 2S 0525 and 
05 5S 0414.
The \ros\, project is supported by the German Bundes\-mini\-ste\-rium f\"ur
Bildung, Wissenschaft, Forschung und Technologie (BMBF/DARA) and the
Max-Planck-Society. This research has made use of the Simbad database, 
operated at CDS, Strasbourg, France.

{}

\bsp
\label{lastpage}
\end{document}